\documentclass[aps,pre,twocolumn]{revtex4-1}
\usepackage{amsmath,bm,epsfig,,subfigure}
\usepackage{color}
\usepackage[normalem]{ulem}
\def\Fbox#1{\vskip1ex\hbox to 8.5cm{\hfil\fboxsep0.3cm\fbox{%
  \parbox{8.0cm}{#1}}\hfil}\vskip1ex\noindent}  %%  {TEXT} in BOX
\newcommand{\Hes}{\B{\C H}}

\newcommand{\B}[1]{{\bm{#1}}}%% Bold Roman & Greek Lower & Upper Case
\newcommand{\C}[1]{{\mathcal{#1}}}    %%   Calligrapfic Upper case
\let \= \equiv \let\*\cdot \let\~\widetilde \let\-\overline

\begin{document}
\title{The Anatomy of Plastic Events in Magnetic Amorphous Solids}
\author{H. George E. Hentschel, Itamar Procaccia and Bhaskar Sen Gupta}
\affiliation{Department of Chemical Physics, The Weizmann Institute of Science, Rehovot 76100, Israel}
\date{\today}
\begin{abstract}
Plastic events in amorphous solids can be much more than just ``shear transformation zones" when the positional degrees
of freedom are coupled non-trivially to other degrees of freedom. Here we consider magnetic amorphous solids where mechanical and magnetic degrees of freedom interact, leading to rather complex plastic events whose nature must be disentangled. In this paper we uncover the anatomy of the various contributions to some typical plastic events. These plastic events are seen as Barkhausen Noise or other ``serrated noises". Using theoretical considerations we explain
the observed statistics of the various contributions to the considered plastic events. The richness of contributions and their
different characteristics imply that in general the statistics of these ``serrated noises" cannot be universal, but rather
highly dependent on the state of the system and on its microscopic interactions.
\end{abstract}
\maketitle

\section{Introduction}

Modeling the mechanical properties of amorphous solids is an active subject of current research, requiring
detailed understanding of the many-body processes that occur in such system when subjected to external strains. External strains can be mechanical, magnetic or electric, depending on the properties of the
amorphous solid in question. The responses of amorphous solids to such external strains is usually not smooth, giving rise
to "serrated" plots of stress vs. strain, energy vs. strain, magnetization vs. external magnetic field etc. A lot of effort was spent on characterizing the probability distribution functions of such serrated responses.
In the context
of magnetic jumps this is referred to as Barkhausen Noise \cite{19Bar,96SBMS,09DMW,95PDS,01SDM,06DZ} but other
serrated noises were studied as well \cite{14SWDS}. In a number of cases strong claims of universality were made.

Recently we have analyzed in some detail model amorphous solids in which
there is a significant coupling between mechanical and magnetic properties \cite{12HIP,13DHPS,13HPS,14HIPS}. Doing so we realized that the characterization of the physics of plastic events can be quite demanding; there is more in these events
than what meets the eye at first impression. The aim of this paper is to highlight the somewhat complex
anatomy of plastic events in such systems, with a word of caution to researchers in the field that similar
complexity may arise in other systems as well, and reasonable modeling should take this into account. In particular
we will conclude below that in general one should not expect universal probability distribution functions since
the statistics of the serrated responses depend on many details of the microscopic interactions and on the state
of the system.

\begin{figure}
\includegraphics[scale = 0.35]{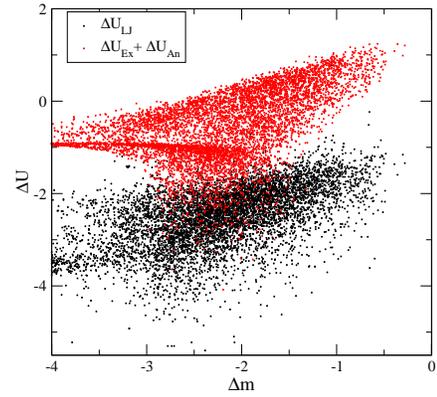}
\caption{Scatter plot of value of the energy drops $\Delta U$ as a function of the
simultaneous changes $\Delta m$ in the magnetization. These occur when the magnetic field
is ramped up and down to form the hysteresis loop of the Barkhausen Noise.}
\label{delu}
\end{figure}

Deferring all details to the next section, we motivate the present paper by showing in Fig. \ref{delu} a scatter plot of the
values of energy drops $\Delta U$ during plastic events when the system is strained by an external magnetic field. The scatter plot is shown as a function of the magnetization jump $\Delta m$ that occurs simultaneously with the energy drop. First, one sees that for a given $\Delta m$ one has a wide
distribution of $\Delta U$ values. Second, these values of the energy drops fall in different groups, with a strange intense line and two triangular groups that are only partly overlapping. Understanding such scatter plots and their implications on the physics of the solid is what we mean by ``the anatomy of plastic events". A theory of plasticity in such system should include also the understanding
of the statistical distribution of such events. The density of points in every little box of size $d\Delta U d\Delta m$ inFig. \ref{delu} is proportional to the joint probability $P(\Delta U,\Delta m)d\Delta U d\Delta m$. Understanding how this probability distribution function is determined by the different physical process requires a theory of the anatomy of plastic events. The aim of this paper
is to provide such a theory for one particular model of amorphous magnetic glass. Other models will require a similar approach.
%%%%%%%%%%%%%%%%%%%%%%%%%%%%%%%%%%%%%%%%

In Sect. \ref{model} we present the model that was introduced recently and analyzed for some of its
aspects in Refs.~\cite{12HIP,13DHPS,13HPS,14HIPS}. The next section \ref{plast} deals with the notion of plastic events, both under
external mechanical strain and under external magnetic field. The anatomy of the plastic events that occur under magnetic straining is
studied in section \ref{anat}, in which we provide the numerical analysis, a theory, and a comparison between the two.  In Sect.  \ref{mechstrain} we discuss briefly the anatomy of plastic events under mechanical straining, including the implications to magnetostriction.
The last section offers a summary and a discussion.

\section{ A Model of a Magnetic Amorphous Solid with Strong Local Anisotropy}
\label{model}

The model Hamiltonian was introduced in \cite{12HIP} and analyzed further in \cite{13DHPS,13HPS,14HIPS,14GSP}. The magnetic part of the model is in the spirit of the Harris, Plischke and Zuckerman (HPZ) Hamiltonian \cite{73HPZ}
but with a number of important modifications. These modification were made to bring the model closer
 to the physics of amorphous magnetic solids \cite{12HIP}. The first major difference is that the particles in the present case are not pinned to a lattice.  We write the Hamiltonian as
\begin{equation}
\label{umech}
U(\{\B r_i\},\{\B S_i\}) = U_{\rm mech}(\{\B r_i\}) + U_{\rm mag}(\{\B r_i\},\{\B S_i\})\ ,
\end{equation}
where $\{\B r_i\}_{i=1}^N$ are the 2-dimensional positions of $N$ particles in an area $L^2$ and $\B S_i$ are spin variables. The mechanical part $U_{\rm mech}$ is chosen to represent a glassy material with a binary mixture of 65\% particles A and 35\% particles B,
with Lennard-Jones potentials having a minimum at positions $\sigma_{AA}=1.17557$, $\sigma_{AB}=1.0$ and $\sigma_{BB}=0.618034$ for the corresponding interacting particles \cite{09BSPK}. These values are chosen to guarantee good glass formation and avoidance of crystallization. The energy parameters chosen are $\epsilon_{AA}=\epsilon_{BB}=0.5$
$\epsilon_{AB}=1.0$, in units for which the Boltzmann constant equals unity. All the potentials are truncated at distance 2.5$\sigma$ with two continuous derivatives. $N_A$ particles A carry spins $\B S_i$; the $N_B$ B particles are not magnetic. Of course $N_A+N_B= N$. We choose the spins $\B S_i$ to be classical $xy$ spins; the orientation of each spin is then given by an angle $\phi_i$ with respect to the direction of the external magnetic field which is along the $x$ axis.

The magnetic part of the potential energy takes the form \cite{12HIP}:
\begin{eqnarray}
&&U_{\rm mag}(\{\B r_i\}, \{\B S_i\}) = - \sum_{<ij>}J(r_{ij}) \cos{(\phi_i-\phi_j)}\nonumber\\&&-  \sum_i K_i\cos^2{(\phi_i-\theta_i(\{\B r_i\}))}-  \mu_A B \sum_i \cos{(\phi_i)} \ .
\label{magU}
\end{eqnarray}
Here $r_{ij}\equiv |\B r_i-\B r_j|$ and the sums are only over the A particles that carry spins. Notice
that in the present model the exchange parameter $J(\B r_{ij})$ is a function of a changing inter-particle position (either due to affine motions induced
by an external strain or an external magnetic field or due to non-affine particle displacements, and see below). Thus randomness in the exchange interaction is coming from the random positions $\{\B r_i\}$, whereas the function $J(\B r_{ij})$ is not random. We choose the monotonically decreasing form $J(x) =J_0 f(x)$ where $f(x) \equiv \exp(-x^2/0.28)+H_0+H_2 x^2+H_4 x^4 $ with
$H_0=-5.51\times 10^{-8}\ ,H_2=1.68 \times 10^{-8}\ , H_4=-1.29 \times 10^{-9}$.
This choice cuts off $J(x)$ at $x=2.5$ with two smooth derivatives.  Note that we need to have at least two smooth derivatives in order to compute the Hessian matrix below. Finally, in our case $J_0=3$.

Another major difference with the HPZ model is that in the present case
the local axis of anisotropy $\theta_i$ is {\em not} selected randomly, but is determined by the local structure. Recall that in a crystalline solid the easy axis is determined by the symmetries of the lattice. In an amorphous solid the arrangement of particles changes from one position to the other, and we need to find the local easy axis by taking this local structure into account. Define  the matrix $\B T_i$:
\begin{equation}
T_i^{\alpha\beta} \equiv \sum_j J( r_{ij})  r_{ij}^\alpha r_{ij}^\beta/\sum_j J( r_{ij}) \ .
\end{equation}
Note that we sum over all the particles that are within the range of $J( r_{ij})$; this is sufficient to take into account the arrangement of the local neighborhood of the $i$th particle. The matrix $\B T_i$ has two eigenvalues in 2-dimensions that we denote as $\kappa_{i,1}$ and $\kappa_{i,2}$, $\kappa_{i,1}\ge \kappa_{i,2}$. The eigenvector that belongs to the larger eigenvalue $\kappa_{i,1}$ is denoted by $\hat {\B n}$. The easy axis of anisotropy is given by $\theta_i\equiv \sin^{-1} (|\hat n_y|)$. Finally the coefficient $K_i$ which now changes from particle to particle is defined as
\begin{equation}
\label{KK}
K_i \equiv \tilde C[\sum_j J( r_{ij})]^2 (\kappa_{i,1}-\kappa_{i,2})^2\ ,~~ \tilde C= K_0/J_0\sigma^4_{AB} \ .
\end{equation}
The parameter $K_0$ determines the strength of this random local anisotropy term compared to other terms in the Hamiltonian. For most of the data shown below we chose $K_0=5.0$. The form given by Eq.~(\ref{KK}) ensures that for an isotropic distribution of particles $K_i=0$. Due to the glassy random nature of our material the direction $\theta_i$ is random. In fact we will assume below (as can be easily tested in the numerical simulations) that the angles $\theta_i$ are distributed randomly in the interval $[-\pi,\pi]$. It is important to note that ramping the magnetic field does NOT change this flat distribution and we will assert that the probability distribution $P(\theta_i)$ can be simply taken as
\begin{equation}
P(\theta_i)d\theta_i = \frac{d\theta_i}{2\pi} \ .
\label{ptheta}
\end{equation}
We have checked in the numerical simulations that Eq. (\ref{ptheta}) is valid to a high approximation
at all values of $B$.
The last term in Eq. (\ref{magU}) is
the interaction with the external field $B$. We have chosen $\mu_A B$ in the range [-0.08,0.08]. At the two extreme values all the spins are aligned along the direction of $\B B$.

In passing we should comment on the chosen parameters in the model. Our guiding line was to choose
parameters such that the magnetostriction coefficient is of the order of what is known in laboratory
materials. Decreasing $K_0$ results in much smaller magnetostriction coefficients and vice versa.
We did not aim at modeling a particular material, and our interest here, as before, is in the generic
properties of amorphous solids with strong local anisotropy.
%%%%%%%%%%%%%%%%%%%%%%%%%%%%%%%%%%%%%%%%%%%%%%%%%

\section{What is Plasticity?}
\label{plast}

It is well known that plasticity in crystalline solids is carried by defects, like dislocations, whose glide under external strains is dissipative, leading to energy loss. What are the mechanisms of energy loss
in amorphous solids is less well known, although research in the last two decades has shed considerable
light on the fundamental physics of plasticity in amorphous solids \cite{99ML,02TWLB,04ML,10KLP,11HKLP,09LP}.
In the present system we can have
two distinct external agents that can strain the system, i.e. mechanical strain and magnetic field.
Such external strain can be studied in systems having finite or zero temperature. Since we are interested
in the anatomy of plastic events we opt for the latter, temperature fluctuations tend to mask the
clear cut plastic events that are recognized at $T=0$. To keep the system at $T=0$ we must also ramp
the external strain or the magnetic field quasistatically to allow the system to remain in mechanical
equilibrium at all times, without heating effects. In such conditions it is completely clear what
are the instabilities that are responsible to plastic events.

\begin{figure}
\includegraphics[scale = 0.35]{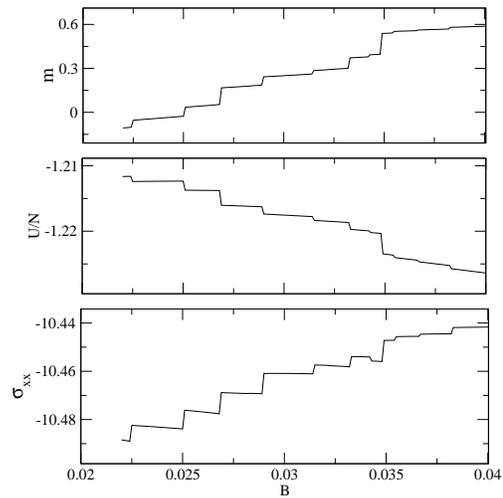}
\caption{The magnetization, the energy per particle and the stress component $\sigma_{xx}$ as a function of external magnetic field $B$. This figure demonstrates that all three quantities have discontinuities at the same value of $B$ where the system undergoes a plastic event with one of the eigenvalues of the Hessian matrix $\Hes$ hits zero, cf. the next figure.}
\label{changes}
\end{figure}

The response of our system to external strain, be it mechanical or magnetic, is reversible and smooth as long as the system is mechanically and magnetically stable. This is the case as long as the Hessian matrix $\Hes$ has only positive eigenvalues. In the present case $\Hes$ takes on the form \cite{12HIP}:
\begin{equation}
\label{Hesa}
\Hes =
\begin{pmatrix}
  \frac{\partial^2U}{\partial \B r_i\partial \B r_j} & \frac{\partial^2U}{\partial \B r_i\partial \phi_j} \\
  \frac{\partial^2U}{\partial \phi_i\partial \B r_i}  & \frac{\partial^2U}{\partial \phi_i\partial \phi_j} \
\end{pmatrix} \ .
\end{equation}
The system loses stability when at least one of the eigenvalues of $\Hes$ goes to zero. When this happens, there appears an instability that results in a discontinues change in stress, in energy and in magnetization. In Fig. \ref{changes} we show a typical blown up section of the energy, magnetization and stress curves as a function of $B$.
%%%%%%%%%%%%%%%%%%%%%%%%%%%%%%%%%%%%%%%%%
We see that the discontinuities appear simultaneously in all the three quantities at the same values of $B$. These are irreversible plastic events that take the system from one minimum in the energy landscape through a saddle-node bifurcation to another minimum in the energy landscape where again all the eigenvalues of $\Hes$ are positive. In Ref. \cite{12HIP} we derived an exact equation for the dependence of any eigenvalue $\lambda_k$ on $B$ for a fixed external strain, which reads:
\begin{equation}
\frac{\partial \lambda_k}{\partial B}{\bf |}_{\gamma}   =   c^{(b)}_{kk} - \sum_\ell \frac{a^{(b)}_\ell [b^{(r)}_{kk\ell}+b^{(\phi )}_{kk\ell}]}{\lambda_\ell} .
\label{diff}
\end{equation}
The precise definition of all the coefficients is given explicitly in Ref. \cite{12HIP}. Generically, when one eigenvalue, say $\lambda_P$ approaches zero, all the other terms in Eq. (\ref{diff}) remain bounded, leading to the approximate equation
\begin{equation}
\frac{\partial \lambda_P}{\partial B}{\bf |}_{\gamma}\approx \frac{\text{Const.}}{\lambda_P} \ . \label{sqsing}
\end{equation}
In such generic situations the eigenvalue is expected to vanish following a square-root singularity, $\lambda_P \sim (B_p-B)^{1/2}$ where $B_p$ is the value of the external magnetic field where the eigenvalue vanishes. The reader should be aware of the fact that at some special values of $B$ it may happen that the coefficient Const in Eq. \ref{sqsing} vanishes at the
instability leading to a an exponent different from 1/2 \cite{13DHPS}. This non generic feature hardly changes the considerations of the present paper.
%%%%%%%%%%%%%%%%%%%%%%%%%%%%%%%%%%%%%%%%%%%%%%%%%%%%%%%%
\begin{figure}
\includegraphics[scale = 0.35]{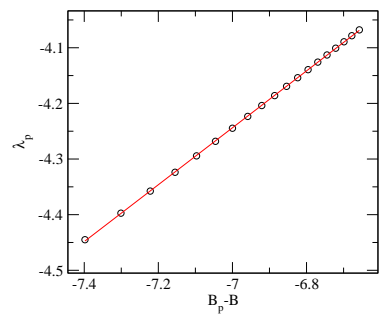}
\caption{The logarithm of the eigenvalue $\lambda_P$ that hits zero at $B_P$ as a function of the logarithm of $B_P-B$. The slope has a value of 1/2. }
\label{instab}
\end{figure}
%%%%%%%%%%%%%%%%%%%%%%%%%%%%%%%%%%%%%%%%%%%%%%%%%%%%%%
In Fig. \ref{instab} we show a typical dependence of the eigenvalue $\lambda_P$ on $B$, where the square-root singularity is apparent. It is also interesting to examine what happens to the eigenfunctions $\B \Psi^{k}$ which are associated with the eigenvalues $\lambda_k$ as the instability is approached. The answer is that all the eigenfunctions of $\Hes$ are delocalized far from the instability, but the one eigenfunction $\B \Psi^{P}$ associated with $\lambda_P\to 0$ gets localized on $n\ll N$ particles. A typical projection of $\B \Psi^{P}$ close to the instability on the particles positions and on the spins is shown in the two panels of Fig. \ref{proj}.
%%%%%%%%%%%%%%%%%%%%%%%%%%%%%%%%%%%%%%%%%%%%%%%%%%%%%%%%
\begin{figure}
\includegraphics[scale = 0.35]{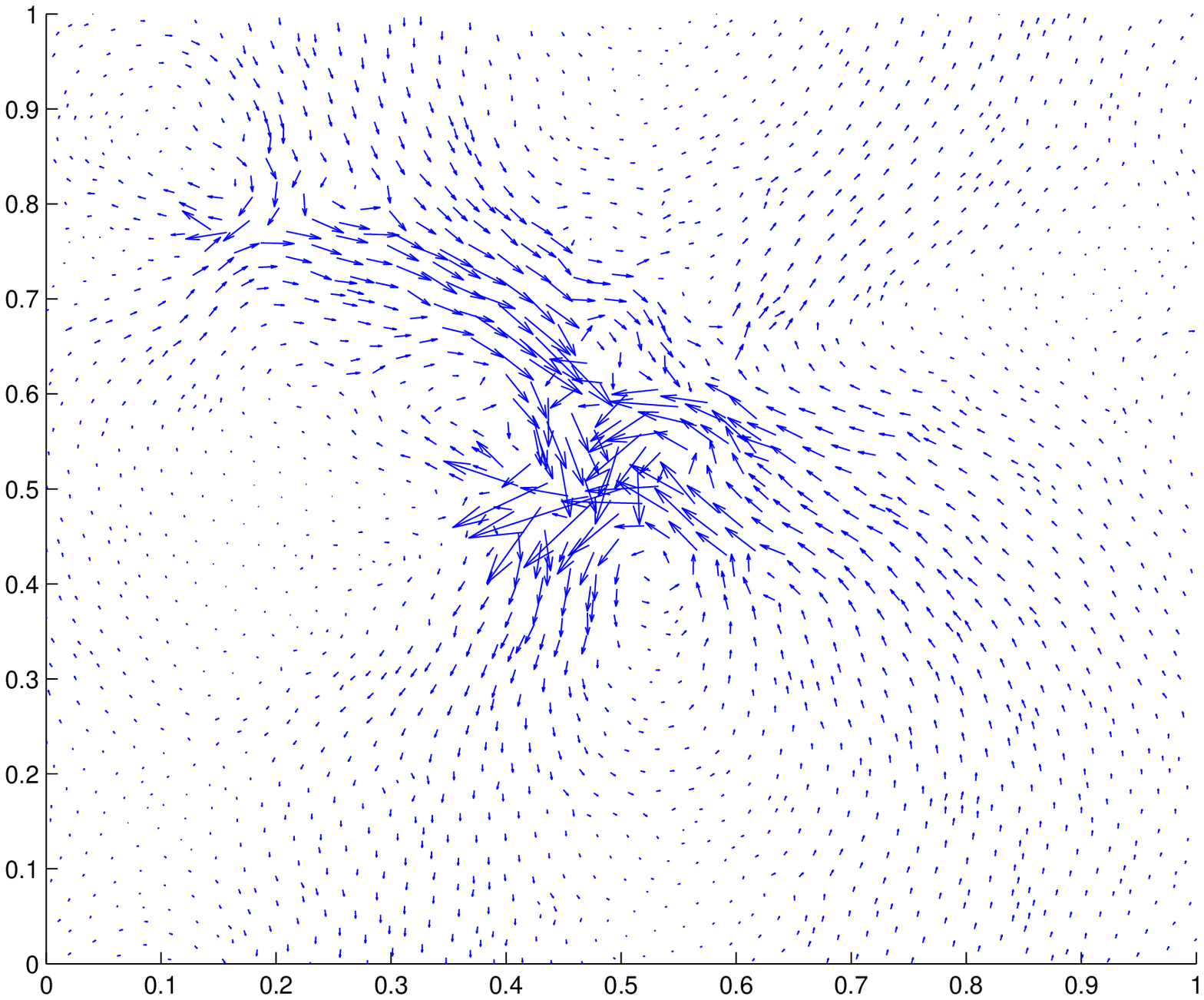}
\includegraphics[scale = 0.35]{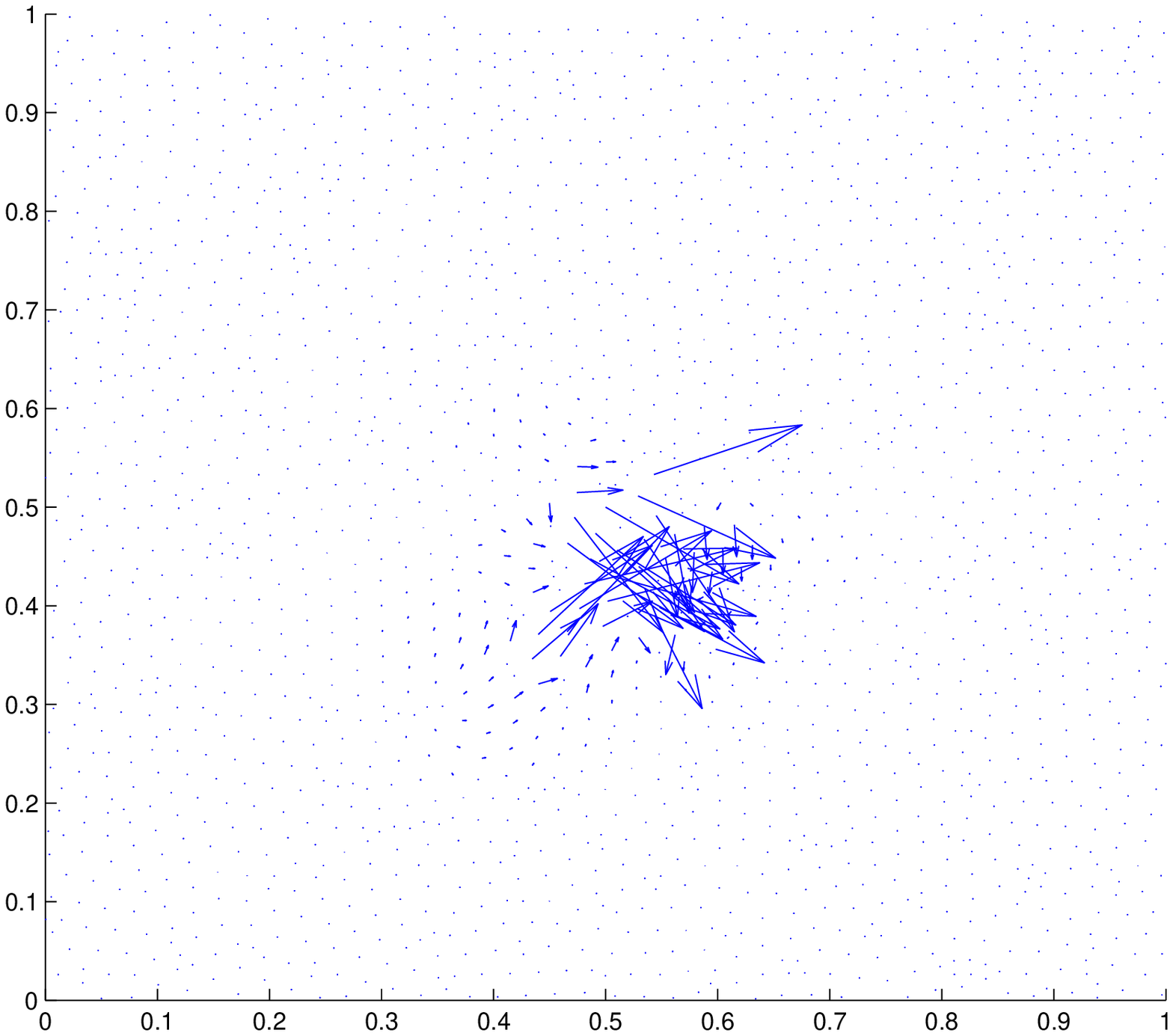}
\caption{The projection of the eigenfunction $\B \Psi^{P}$ associated with the eigenvalue $\lambda_P$ shown in Fig. \ref{instab} projected on the particles positions and on the spins in the upper and lower panels respectively. The upper panel shows a typical non-affine displacement field associated with a plastic event, having the quadrupolar structure of an Eshelby solution. The lower panel shows that the same event is associated with a co-local flip of spins, leading to the change $\delta m$ of the Barkhausen Noise.}
\label{proj}
\end{figure}
%%%%%%%%%%%%%%%%%%%%%%%%%%%%%%%%%%%%%%%%%%%%%%%
We see that the non-affine movement of the particles is very similar to the standard ``Eshelby like" quadrupolar event that is so typical to amorphous solids. The projection on the spin degrees of freedom shows that a patch of spins had changed its orientation (magnetic flip of a domain). Note that the patch is compact, without any fractal or other esoteric characteristics that were associated with Barkhausen Noise in the past.  This is the nature of the event that is associated with the Barkhausen Noise in our case. The reader should
be aware however of the fact that the addition of long range dipole-dipole interactions can change this qualitatively, leading
to elongated magnetic domains and a different mechanism of Barkhausen Noise due to the movement of domain boundaries \cite{06DZ}.

\section{The anatomy of plasticity}
\label{anat}

\begin{figure}
\includegraphics[scale = 0.35]{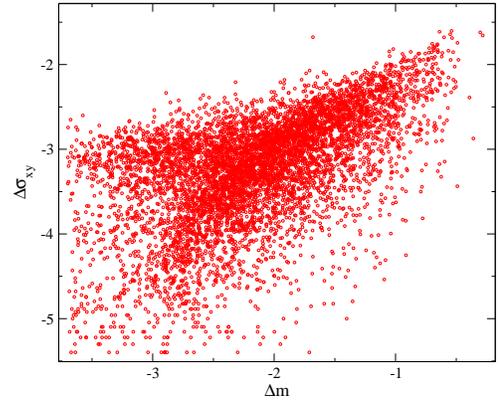}
\caption{Scatter plot of value of the stress drops $\Delta \sigma$ as a function of the
simultaneous changes $\Delta m$ in the magnetization. These occur when the magnetic field
is ramped up and down to form the hysteresis loop of the Barkhausen Noise.}
\label{delsig}
\end{figure}

The interesting physics of plasticity in this model stems from the fact that the Hessian matrix
(\ref{Hesa}) couples the positional to the magnetic degrees of freedom. Thus a plastic drop in stress and
energy will be usually coupled also to a change in the magnetization. Whether one strains the system
with a mechanical strain or a magnetic field, the plastic drops will be composite processes in which
all the degrees of freedom contribute to the non-affine response. In this section we focus on these
events that are triggered by the magnetic field as the straining agent. To expose the anatomy of the plastic events
we re-plot the data in Fig. \ref{changes} in a different way, i.e. as a scatter plot of the drops in
energy or in stress as a function of the magnetization change $\Delta m$.  The first was shown in
Fig. \ref{delu} and the second is shown here as Fig. \ref{delsig}.
%%%%%%%%%%%%%%%%%%%%%%%%%%%%%%%%%%%%%%%%
%%%%%%%%%%%%%%%%%%%%%%%%%%%%%%%%%%%%%%%%%%%%%%%%%%%%%%%%%%%%%%%%%%%%%%%
As before, we see that also the values of the stress drops organize into two distinct groups that are however
not non-overlapping.
%%%%%%%%%%%%%%%%%%%%%%%%%%%%%%%%%%%%%%%%%%%
\subsection{Detailed analysis of the energy drops}
\label{endrop}
\begin{figure}
\includegraphics[scale = 0.35]{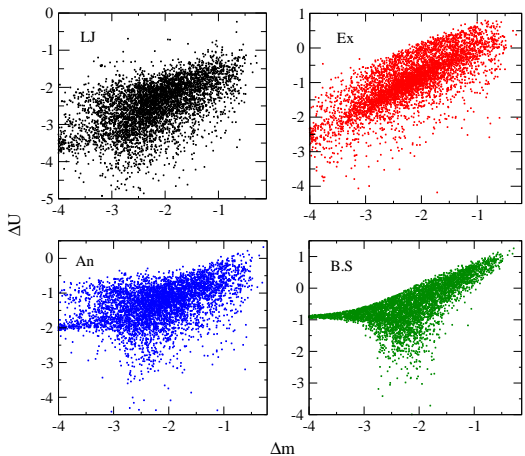}
\caption{The four distinct contribution to energy drops in plastic events. The combination of all these scatter plots should yield
the data in Fig. \ref{delu}}.
\label{anatdelu}
\end{figure}

Our first task is to rationalize the distributions that appears in figures like Figs.~\ref{delu} and \ref{delsig}. Focussing
as an example on the energy drops,
we return to the Hamiltonian and find out which of the terms is responsible to which group of energy drop values
in these figures. This separation is demonstrated in Fig. \ref{anatdelu},
where the energy drop is assigned to four different contributions to the Hamiltonian, i.e. the
Lennard-Jones positional degrees of freedom, the exchange interaction, the anisotropy energy and finally the
interaction with the magnetic field.
Obviously the combination of the scatter plots in Fig. \ref{anatdelu} will lead to what was shown
as Fig. \ref{delu}. The same decomposition can be done for the stress drops but for the sake of brevity we focus here on understanding
the results shown in Fig. \ref{anatdelu}; a similar analysis for the stress drops is implied.
%%%%%%%%%%%%%%%%%%%%%%%%%%%%%%%%%%%%%%%%%%%%%%%%%%%%%%%%%%%%%

To understand what we see we will invoke the result of the previous work \cite{14HIPS} in which the distribution $P(\Delta m)$ was measured over three orders of magnitude $10^{-3.5}<\Delta m< 10^{-.5}$. These magnetization changes involved flips of magnetic domains of between $10^{0}<\Delta n< 10^{3}$ particles. This probability distribution function (PDF) was found to be well fitted in this regime by a form
\begin{equation}
\label{final}
P(\Delta m)=\frac{\exp(-A \Delta m ) }{\Delta m} f(\Delta m) \ ,
\end{equation}
where the exponential decay rate $A$ is analytically computed and the function $f(\Delta m)$ is evaluated explicitly.

To make the connection to our present data we need to discuss the conditional distribution $P_2(\Delta U|\Delta m)$ in terms of which the joint distribution
\begin{equation}
\label{P}
P(\Delta U, \Delta m)= P(\Delta U|\Delta m) P(\Delta m)
\end{equation}
can be written.

\noindent We shall start with the general form for the energy drop
\begin{equation}
\label{U}
\Delta U = \sum_{i=1}^{n}u_i ,
\end{equation}
where $u_i$ is the energy change (both mechanical and magnetic) associated with the slip of the $i$th spin in the flipping domain and $n $ is the number of flipping spins. We assume that each spin flip contributes to $\Delta m$ and thus
\begin{equation}
n\sim C \Delta m\ , \label{nsimM}
\end{equation}
with some unknown constant $C$.
Thus $\Delta U$ is a sum of $n$ random variables, and using the central limit theorem we can assume a Gaussian form for the conditional probability
\begin{equation}
\label{P2}
P_2(\Delta U|\Delta m) = \frac{1}{\sqrt{2\pi \sigma^2}}\exp{-(\Delta U - \langle\Delta U\rangle )^2/(2\sigma^2)}.
\end{equation}
where both the average energy drop $\langle\Delta U\rangle = \langle\Delta U\rangle (\Delta m)$ and the variance of the energy drops $\sigma^2 = \sigma(\Delta m)^2$ are functions of $\Delta m$.

\noindent We can now introduce two exponents $\zeta_1$ and $\zeta_2$ by
\begin{eqnarray}
\label{zeta}
\langle\Delta U\rangle (\Delta m) &=& K_1 \Delta m^{\zeta_1} \nonumber \\
\sigma^2(\Delta m) &=& K_2 \Delta m^{\zeta_2}.
\end{eqnarray}
One of the purposes of these notes is to estimate these exponents $\zeta_1$ and $\zeta_2$.

Now  the total energy drop can be separated into its individual mechanical and magnetic contributions
\begin{eqnarray}
\label{energy}
\Delta U  & = & \Delta U^{mech}+ \Delta U^{mag} \nonumber \\
               & = & \Delta U^{mech}+ \Delta U^{ex}+\Delta U^{anis}+\Delta U^{b}.
\end{eqnarray}
where $\Delta U^{mech}$ is the mechanical contribution to the energy drop, while the spin contribution can be separated into exchange $\Delta U^{ex}$, anisotropic $\Delta U^{anis}$ and magnetic field $\Delta U^{b}$ contributions, and depending on the spin Hamiltonian for the metallic glass (which in our case is Eq. (\ref{magU})) may lead to different exponents.

Let us therefore analyse the consequences of Eq. (\ref{U}) carefully. First we note that
\begin{equation}
\label{Uav}
\langle\Delta U\rangle (\Delta m) = \sum_{i=1}^{n}\langle u_i \rangle ,
\end{equation}
and thus if a non zero $ \langle u_i \rangle=\langle u\rangle$ exists we would find that $\langle\Delta U\rangle (\Delta m) \sim \Delta m$ or $\zeta_1=1$. On the other hand if there exists a contribution consisting of $n$ random variables with zero mean, then we might expect that contribution to scale like $ \sim \Delta m^{1/2}$ or $\zeta_1 = 1/2$. The important point to note is that by measuring $\zeta_1$ the physics underlying the spin flips can be found.

To estimate the variance of the fluctuations $\sigma^2 = \langle\Delta U^2\rangle - \langle\Delta U\rangle^2$  we write using Eq.~(\ref{U})
\begin{equation}
\label{var}
\sigma(\Delta m)^2 =  \sum_{i=1}^{n}\sum_{j=1}^{n}[\langle u_i u_j\rangle -\langle u_i \rangle \langle u_j \rangle].
\end{equation}
Thus while $\langle\Delta U\rangle$ only depends on the additive contribution of $n$ random variables,  the variance also depends on the correlation of different spins within the flipped domain.
To see the consequence of these correlations, let us consider first the case where the spins $i$ and $j$ are uncorrelated. In that case $ \sigma(\Delta m)^2 =  n [\langle u^2\rangle -\langle u\rangle^2] \sim \Delta m$ or $\zeta_2 = 1$.  On the other hand, in the limit of strong correlations between spins in the flipped domain  $[\langle u_i u_j\rangle -\langle u_i \rangle \langle u_j \rangle] \neq 0$ and as a consequence $\sigma(\Delta m)^2 \sim \Delta m^2$ or $\zeta_2 = 2$.  It is also possible that the clean scaling described above may not exist but rather several mechanisms are mixed. Only simulations and data analysis can answer these questions.

At this point we want to use the results of Ref. \cite{14HIPS} in which an analytic form for the probability
to see a magnetic jump of $\Delta m$ was proposed. To this aim we will attempt to find analytic approximants
to the pdf $P(\Delta U|\Delta m)$ in the form of four distinct contributions in agreement with Fig. \ref{anatdelu}:
\begin{eqnarray}
&&P(\Delta U|\Delta m) = P_{LJ}(\Delta U|\Delta m)+P_{Ex}(\Delta U|\Delta m)\nonumber\\
&&+P_{An}(\Delta U|\Delta m)+P_{B\cdot S}(\Delta U|\Delta m)\ . \label{4cont}
\end{eqnarray}
Analyzing the data shown in Fig. \ref{anatdelu} we could approximate each of the four conditional pdfs
in the form
\begin{equation}
P_i(\Delta U|\Delta m) \approx e^{-\frac{\left(\Delta U-K_1(\Delta m)^{\zeta_1}\right)^2}{K_2(\Delta m)^{\zeta_2}}}. \label{gauss}
\end{equation}
The best fits for the exponents $\zeta_1$ and $\zeta_2$ and for the constant $\sigma$ are provided in table \ref{tab}.

\begin{center}
\begin{table}
\begin{tabular}{ |c|c|c|c|c| }
 \hline
 & LJ & Ex & An & B.S \\
 \hline
 $\zeta_1$ & 0.5 & 1.0 & 0.5 & 1.0 \\
 $\zeta_2$ & 0.75 & 1.0 & 1.0 & 1.0 \\
 \hline
\end{tabular}
\caption{The best fit values of $\zeta_1$ and $\zeta_2$ for different energy terms. Note that the exponent of 0.75 is not
explained theoretically but the data for this particular contribution is small, close to the noise level.}
\label{tab}
\end{table}
\end{center}

\begin{figure}
\includegraphics[scale = 0.42]{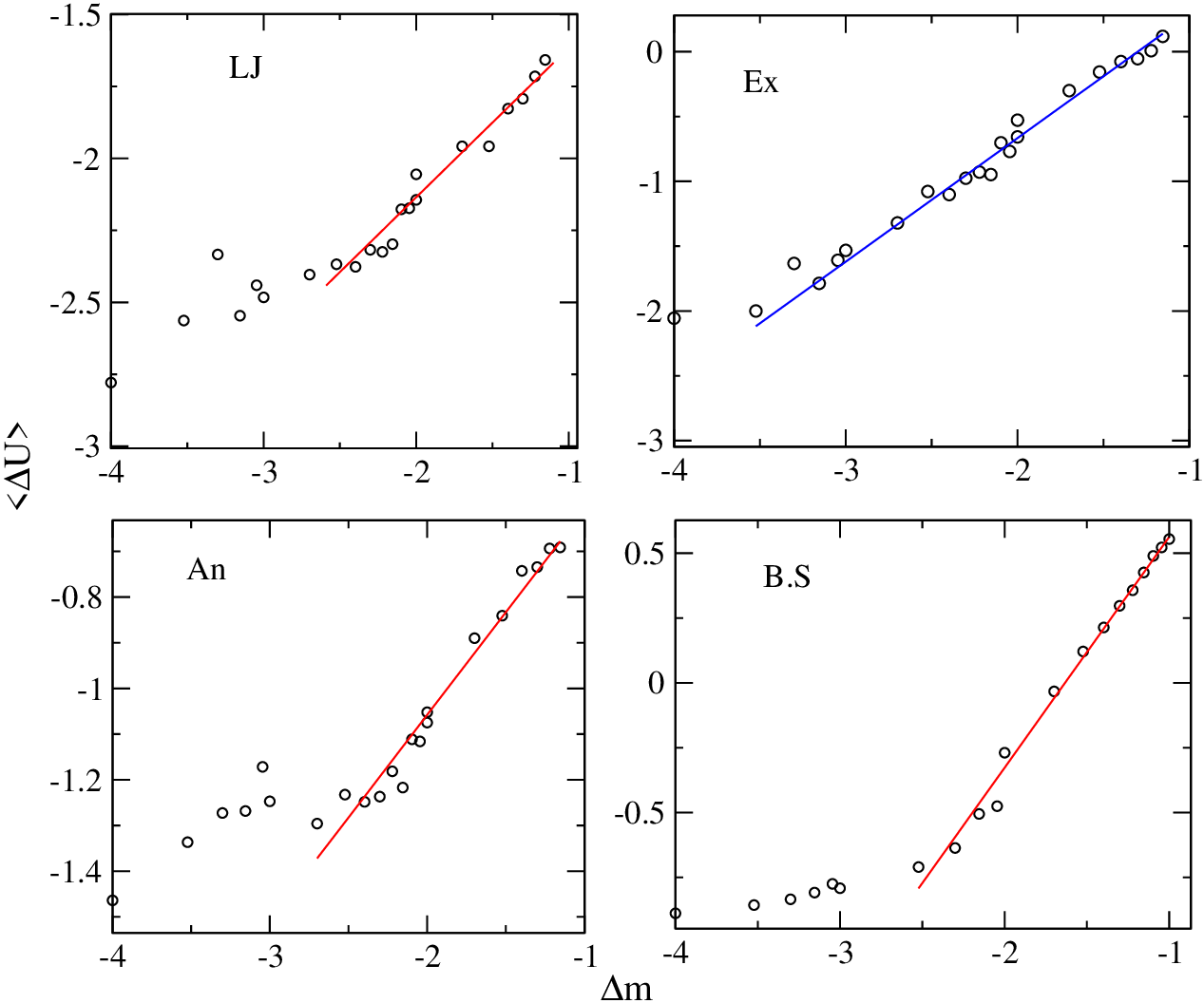}
\caption{The mean of $\langle \Delta U\rangle$ as function of $\Delta m$ for the four different
contributions that play a role in our system.}
\label{mean}
\end{figure}

The quality of the fits are demonstrated in Figs. \ref{mean} and \ref{variance}. The exponents agree with our expectations
of being 1/2 or 1, except for the variance of the Lennard-Jones contribution in Fig.~\ref{variance}. We note however that the
amplitude of the Lennard-Jones $\Delta U$ is two orders of magnitude less than the other contributions and therefore we
we are close to the noise level and cannot trust this particular measurement. The smallness of this contribution arises from the
fact that the straining here is done magnetically and the position of the particles do not change that much.
%%%%%%%%%%%%%%%%%%%%%%%%%%%%%%%%%%%%%%%%%%%%%%%%%%%%%%%%%%%%%%%%%%%
%%%%%%%%%%%%%%%%%%%%%%%%%%%%%%%%%%%%%%%%%%%%%%%

\begin{figure}
\vskip 1 cm
\includegraphics[scale = 0.40]{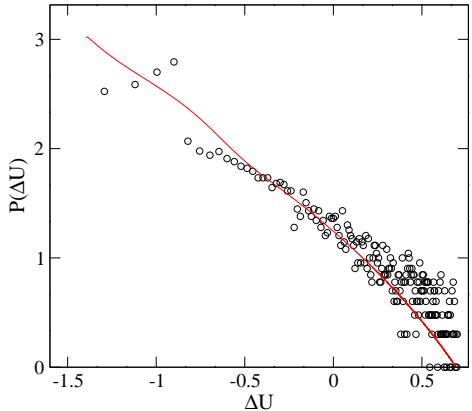}
\caption{Comparison of the direct calculation of $P(\Delta U)$ in plastic events due to straining
with the magnetic field to the reconstruction of the same quantity from the anatomical dissection
of this quantity and the independent knowledge of $P(\Delta m)$ \cite{14HIPS}}
\label{compare}
\end{figure}

Finally we write $P(\Delta U)$ in the form
\begin{eqnarray}
&&P(\Delta U) = \int d\Delta m \Big[P_{LJ}(\Delta U|\Delta m)+P_{Ex}(\Delta U|\Delta m)\nonumber\\
&&+P_{An}(\Delta U|\Delta m)+P_{B\cdot S}(\Delta U|\Delta m)\Big] P(\Delta m)\ .
\label{marginal}
\end{eqnarray}
Now we use the analytic form of $P(\Delta m)$ from Ref. \cite{14HIPS} and compute the integral (\ref{marginal})
numerically. The comparison of this reconstruction of the pdf of $\Delta U$ to its direct numerical calculation
is shown in Fig. \ref{compare}.
\begin{figure}
\includegraphics[scale = 0.65]{anatFig.8.eps}
\caption{The variance in the energy changes as a function of $\Delta m$ for the four different contributions that play a role
in our system.}
\label{variance}
\end{figure}

The conclusion of this exercise is that providing the anatomical details of the plastic events can
help in understanding the statistics of energy or stress drops. We do not repeat in this paper the exercise
for the stress drops since it follows verbatim the same steps.

\section{Straining mechanically}
\label{mechstrain}

\begin{figure}
\includegraphics[scale = 0.35]{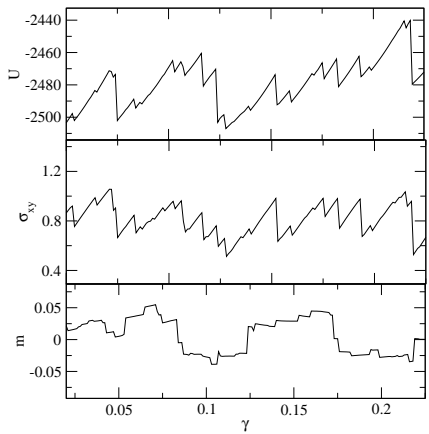}
\caption{Typical dependence of the energy, stress, and magnetization for zero external magnetic field.
Note that all the plastic events occur simultaneously for all quantities. In the present case
the magnetization is fluctuating up and down around a zero mean value.}
\label{mechstress1}
\end{figure}

\begin{figure}
\includegraphics[scale = 0.35]{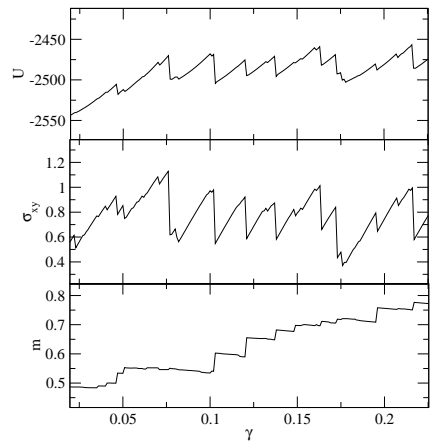}
\caption{Typical dependence of the energy, stress, and magnetization for external magnetic field $B=0.01$.
In the present case the plastic events cause the magnetization to increase; we called this phenomenon
``plasticity induced magnetization" \cite{13HPS}.}
\label{mechstress2}
\end{figure}

A similar richness in the anatomy of plastic events is found when the system is strained mechanically \cite{13HPS}. Even though we strain mechanically the coupling between positional and spin degrees of freedom results again
in having a change in magnetization together with drops in energy and in stress. In Figs.~\ref{mechstress1}
and \ref{mechstress2}
we show a typical plot of energy, stress and magnetization vs external strain, for zero magnetic field
and for a finite magnetic field. Note that in the first case the total magnetization remains zero on the average, with the flips in magnetization $\Delta m$ being negative or positive with equal probability.
For finite magnetic field magnetization is accumulated in the direction of the magnetic field at
each plastic event.
%%%%%%%%%%%%%%%%%%%%%%%%%%%%%%%%%%%%%%%%%%%%%%%%%%%%%%%%%%%%%%%%%%%%%%%%%%%%%%
%%%%%%%%%%%%%%%%%%%%%%%%%%%%%%%%%%%%%%%%%%%%%%%%%%%%%%%%%%%%%%%%%%%%%%%%%%%%%%%%%%%%%%%%
As before with ramping the magnetic field we see that also with mechanical strain the plastic events
couple mechanical and magnetic degrees of freedom. The drops in energy are occurring as a result of plastic
instabilities at the same values of $\gamma$ as the drops in stress and magnetization. The mechanism
is the same, i.e. an eigenvalue of the Hessian matrix hits zero punctuating the smooth curves of energy,
stress or magnetization with sharp drops of irreversible events.

To understand the serrated response curve of the energy or the stress one needs again to search for the
anatomy of the events, displaying carefully the contribution of each physical mechanism for either
energy or stress drop. Since we did not compute independently the 'Barkhausen Noise' $P(\Delta m)$ in this case
we do not repeat the exercise for the case of mechanical straining. We stress however that any interested
researcher must pay attention to the rich physics that is underlying the serrated "noisy" character of the data shown in
Figs. \ref{mechstress1} and \ref{mechstress2}.

\section{Summary and Conclusions}.

The main conclusion of this paper is that {\em characterizing} and {\em understanding} the statistics of
serrated noise is not necessarily the same. Even if we can plot the pdf's of energy drops or of magnetic jumps and
measure the exponent that is associated with their log-log plot, it does not mean that we uncovered the intricate
physics that underlies the phenomenon. We have seen here that even the simplest coupling between mechanical
and magnetic degrees of freedom results in a multitude of contributions to the energy changes upon plastic
events. Each contribution comes with its own statistics, its own exponent and its own amplitude. Of course,
once we have the full information of all the contributions we can reconstruct the pdf of any wanted
quantity, cf. Fig. \ref{compare}. The full information is however not always available in experimental
systems. Thus great care is called for interpreting the observed statistics of serrated noises. In particular
we should stress that changing conditions (like zero or non-zero magnetic field in Figs. \ref{mechstress1} and \ref{mechstress2})
may change the statistics of the serrated noise. The amplitude of the various contributions to the observed serrated
response can depend on the state of the system etc. Thus universal statistics is expected to be the exception rather
than the rule. Rather, a careful analysis of the physics underlying the observed response is called for.

\acknowledgments
This work had been supported in part by an ERC ``ideas" grant STANPAS.

\end{document}